\documentclass[aps,pra,twocolumn,reprint,showpacs,preprintnumbers,superscriptaddress,%
amsmath,amssymb,floatfix,a4paper]{revtex4-1}
\usepackage{epsfig}
\usepackage{epsf}
\usepackage{amsmath}
\usepackage{amsfonts}
\usepackage{amssymb}
\usepackage{bm}
\usepackage{bbm}
\usepackage{nicefrac}
\usepackage{color}
\usepackage{pifont}
\usepackage{graphicx}
\usepackage{epsfig}
\usepackage{latexsym}

\def\d{{\rm d}}
\def\ii{{\rm i}}
\def\e{{\textrm e}}
\def\perm{\zeta}
\def\permdc{\chi}
\def\entr{Q}
\newcommand{\calO}{\mathcal{O}}

\newcommand{\opP}{\mathcal{P}}
\newcommand{\opQ}{\mathcal{Q}}
\newcommand{\paper}{paper}

\newcommand{\colol}{(Color online) }
\DeclareMathOperator{\tr}{Tr}
\def\vecb{\vec}

\begin{document}

\title{Theoretical study of the Compton effect with correlated three-photon emission:\\
From the differential cross section to high-energy triple-photon entanglement}

\author{Erik L\"otstedt}
\email{lotstedt@chem.s.u-tokyo.ac.jp}
\affiliation{Laser Technology Laboratory, RIKEN Advanced Science Institute, 
2-1 Hirosawa, Wako, Saitama 351-0198, Japan}

\author{Ulrich D. Jentschura}
\affiliation{Department of Physics, Missouri University of Science and Technology, 
Rolla, Missouri 65409-0640, USA}
\affiliation{MTA--DE Particle Physics Research Group, 
P.O.Box 51, H--4001 Debrecen, Hungary}

\begin{abstract}
The three-photon Compton effect is studied.
An incoming photon undergoes triple scattering off  a free
electron, which leads to the emission of three entangled photons.  We investigate the
properties of  both the total cross section, assuming a  low-energy cutoff for
the detected photons, and the differential cross section. Particular emphasis
is laid on evaluating polarization-resolved cross sections. 
The entanglement of the final three-photon  state is analyzed.
\end{abstract}



\maketitle

%
%
\section{Introduction}

A photon colliding with a free electron is one of the most  basic processes of
quantum electrodynamics (QED). At low photon energies $\omega_0$ (in the rest
frame of the electron), the only possible process is the scattering of the
incoming photon off the electron.  The electron-positron pair production
threshold is at $\omega_0 = 4m$, which is larger than $2m$ due to the necessity
of providing for the minimum electron recoil momentum. (Here, $m$ is the mass
of the electron. Throughout this 
article, we work in natural units such that $\hbar=c=\epsilon_0=1$, and
$\alpha=e^2/4\pi$, where $\alpha\approx 1/137.036$ is the fine-structure
constant and $e$ the charge of the electron.)  Moreover, for $\omega_0 \ll m$,
the scattering is elastic and referred to as Thomson scattering.  In this
limit, the  Klein-Nishina cross section~\cite{KlNi1929} calculated from
quantum electrodynamics (QED) agrees with the prediction from classical
electrodynamics.  When $\omega_0$ becomes comparable to $m$ or above, the
scattering process  is termed Compton scattering, or just the Compton
effect~\cite{Co1923}. It has to be described in fully relativistic
QED~\cite{JaRo1980,ItZu2006}. 

Compton scattering has been widely studied, and has a large number of
applications. By analyzing the broadening of the Compton peak (the Compton
profile) of the scattered photons, information on the electron momentum
distribution in atoms \cite{DuCo1945,AS2005}, molecules \cite{HaNyVaItSaHa2009}
and condensed matter \cite{KoNaKaSaHiMu2004,BeBaKloetal2011,OlTiLaHaHuHo2012}
can be obtained.  Compton scattering from bound electrons can in general be
described by the Klein-Nishina cross section if the energy gained by the
electron is much larger than the binding energy, which implies that the
electron can be regarded as free during the collision.  It is also possible to
produce high-energy gamma photons through Compton backscattering of laser
photons off of energetic electrons from an accelerator.  The (non-exhaustive list
of) review articles~\cite{BlGo1970,Co1985rpp,GlRe2009,WeEtAl2009} discuss
applications of the Compton effect.  There also exists a nonlinear
generalization of the Compton effect: in a laser field, several photons  are
absorbed by an electron to produce one final photon. The electron-laser
interaction has to be taken into account beyond perturbation theory.  This
nonlinear process has received a lot of interest recently, both
theoretically~\cite{HaHeIl2009,MaDPKe2010,SeKe2011} and
experimentally~\cite{BuEtAl1996,BaElAl2006}. 

Much less studied are the processes where a photon collides with an electron
and splits into two or more final-state photons.  Such a reaction is of higher
order in $\alpha$ and has a smaller cross section.  One should not confuse this
kind of scattering process with multiple single Compton scattering events,
which can occur when a photon  scatters consecutively at different electrons
inside a material~\cite{FePa1980}.  This \paper{} exclusively deals with the
process involving one electron and one photon in the initial state, and a final
state consisting of one electron and one, two or three photons.

At moderate
energies $\omega_0\sim m$, the cross sections of the higher-order processes are
suppressed with one factor of $\alpha$ for each additional emitted photon. A
complete QED calculation of the double Compton effect, where two photons are
emitted,  was first presented in~\cite{MaSk1952}, and has since been
verified experimentally by several groups
\cite{Ca1952,MGBrKn1966,SaDeSiGh1999,SaDeSaSiGh2000,%
SaSaSi2006,SaSiSa2008,SaSiSa2011}. 
The total double Compton cross section, for some particular
photon energy infrared cutoff, has been studied theoretically 
by numerical integration~\cite{RaWa1971}.  The
theory of the nonlinear (multi-photon) double Compton effect
(in the background of a strong laser field) has been given 
only recently~\cite{LoJe2009prl,LoJe2009pra,SeKe2012,MaDP2012}.

The next-order Compton process is  the triple, or three-photon 
 Compton effect, where one photon is
split into three after the collision with a free electron.
A rather sophisticated pertinent experiment has been described
in Ref.~\cite{MGBr1968}; otherwise the experimental 
literature on triple scattering 
appears to be scarce. In Ref.~\cite{MGBr1968},
the differential cross section
(averaged over the detector solid angles) was estimated for one 
specific arrangement of the detection geometry of the emitted photons. 
The three detectors were arranged in a symmetric configuration and 
each detector covered a narrow solid angle $\Omega \ll 4 \pi$. On the
theoretical side, the literature also is very scarce. 
In Ref.~\cite{MaMaDh1959},
the total cross section for the $n$-tuple Compton effect was studied 
 for extremely high photon energies 
$\omega_0 \gg m$ (in the rest frame of the electron).
At moderate energies $\omega_0\sim m$, which could more realistically
be achieved in the laboratory, 
we have recently presented calculations of  the total and differential cross
section for a number of examples of experimentally 
realizable parameter sets~\cite{LoJe2012}. 
It is the purpose of the present \paper{} to  
extend the parameter range covered in Ref.~\cite{LoJe2012}
and to give the details of the  method of calculation 
which could be useful if the method is to be 
adapted to a particular experimental geometry in the future.

Compton scattering with multiple photons in the final state is interesting 
for an additional reason: The final photons are quantum mechanically entangled. The experimental 
production of multi-photon entangled states is currently at the focus of intense research efforts
\cite{GrHoShZe1990,KeEtAl1998,%
 PaEtAl2000,EiEtAl2003,GrBo2008,WeOhDu2010,HuEtAl2010,%
AnGeKr2011,CoGPUr2011,DoBoBoBeLe2012}.
One can say that the  three-photon Compton process is the most basic 
QED process that is able to produce a three-photon 
entangled final state. A somewhat related process is
electron-positron annihilation into three photons. This process has been
studied  both in high-energy physics with colliding $\e^+$ and $\e^-$ beams 
\cite{Gu1954,Gu1955_pr1,BeGa1973,BeKl1981,AdEtAl1992,%
ZeDiRo2002} and in the low-energy domain in the 
context of the decay of orthopositronium as a test of CP violation
\cite{BeLoNa1988,ArEtAl1988,%
SkvH1991,VeFr2003,YaNaAsKo2010,%
AbAdYo2011}. Higher-order QED corrections to the decay rate have been
calculated~\cite{OrPo1949,CaLeSa1977,%
AdFeSa2002,KnKoVe2008,AdDrRaFe2010}.
The discrepancy between the experimental results
of the Tokyo group~\cite{AsOrSh1995,JiAsKo2003} and
the Michigan group~\cite{WeGiCoRi1987prl,WeGiCoRi1989pra,NiGiRiZi1990}
was finally resolved in Ref.~\cite{VaZiGi2003}.

We proceed as follows. In Sec.~\ref{Sec:theory}, we
describe the QED theory necessary to obtain expressions for the differential
cross sections for the two-photon and three-photon Compton effect. A numerical
evaluation of the total cross section is presented in Sec.~\ref{subsec:totcrosssection}, 
and examples of the differential cross section are presented in
Secs.~\ref{subec:diffcrosssec} and \ref{subsec:diffcrosssecGeV}. 
The interesting subject of polarization entanglement among the
three final state photons is discussed in Sec.~\ref{Sec:Entanglement}, and we
conclude in Sec.~\ref{Sec:Concl}.

%
%
\section{Theory}
\label{Sec:theory}

%
%
\subsection{Bispinors and photon states}

In the following, we write the scalar product of two four-vectors $a$ and
$b$ as $a\cdot b\equiv a^\mu b_\mu=a^0 b^0-\vecb{a}\cdot \vecb{b}$, which
also defines the metric convention. The contraction of a four-vector $a$  with
the Dirac gamma matrices $\gamma^{\mu}$ is denoted as $\hat{a}=\gamma^\mu
a_\mu=a^0 \gamma^0-\vecb{a}\cdot \vecb{\gamma}$.

The incoming and outgoing four-vectors  of the electron are labeled as
\begin{equation} 
p_{i,f}=(E_{i,f},\vecb{p}_{i,f}) \,,
\end{equation} 
respectively. The electron
bispinors are used in the representation \cite{ItZu2006}
\begin{equation}\label{bispinordef}
u_r(p)=\sqrt{\frac{E+m}{2m}}
\left(\begin{array}{c}
\delta_{r1}\\
\delta_{r2}\\
\dfrac{1}{E+m} \vecb{\sigma}\cdot \vecb{p} 
\left( \begin{array}{c} \delta_{r1} \\ \delta_{r2} \end{array} \right)
\end{array}\right),
\end{equation}
where $\delta_{ij}$ is the Kronecker delta, 
$r=1$ or $2$ labels the spin of the electron, and 
the vector $\vecb{\sigma}$ is 
composed of the (Pauli) $2 \times 2$ spin matrices,
\begin{equation}
\vecb{\sigma}=\left(
\left[\begin{array}{cc}0&1\\1&0\end{array}\right],
\left[\begin{array}{cc}0&-\ii \\\ii & 0\end{array}\right],
\left[\begin{array}{cc}1&0\\0&-1\end{array}\right]
\right).
\end{equation}
With this 
convention, the spinors are normalized according 
to $u_r^\dagger(p)\gamma^0 u_r(p)=\bar{u}_r u_r=1$.
Here, $\bar{u}_r = u^\dagger_r \, \gamma^0$ is the Dirac adjoint.
The gamma matrices are used in the Dirac representation,
\begin{align}
\label{diracgammadef}
& \gamma^0=\left( \begin{array}{cc} 
\mathbbm{1}&0\\
0&-\mathbbm{1} \end{array}\right),
\qquad
\gamma^i = \left( \begin{array}{cc} 0 & \sigma^i\\
-\sigma^i & 0 \end{array}\right),
\end{align}
for $i=1,2,3$, where $\mathbbm{1}$ denotes the $2 \times 2$ unit matrix,
and the $\sigma^i$ are the components of the vector of Pauli matrices.
The propagation wave vectors (four-vectors) of the  photons are denoted as
\begin{equation}
k_j = (\omega_j,\vecb{k}_j) = 
\omega_j (1, \sin\theta_j\cos\phi_j, \sin\theta_j\sin\phi_j,\cos \theta_j),
\end{equation}
where $\phi_j$ measures the azimuth and $\theta_j$ measures the 
polar angle ($j=0,1,2,3$). 
We take $j=0$ to denote the (incoming) absorbed photon, and 
$j=1,2,3$ to denote the emitted photons. We furthermore define 
\begin{equation}
n_j =
\frac{k_j}{|\vecb{k}_j|} = 
(1, \sin\theta_j\cos\phi_j, \sin\theta_j\sin\phi_j,\cos \theta_j),
\end{equation}
so that the four-vector $k_j$ is given as $k_j =\omega_j n_j$.
In all examples presented in Sec.~\ref{examples}, the 
angles and energies of the final particles are measured in the lab frame, 
in a coordinate system with the polar axis defined by the incoming photon, 
i.e., $k_0=\omega_0(1,0,0,1)$. A head-on collision such that 
$p_i=(E_i,0,0,-\sqrt{E_i^2-m^2})$ is always assumed.

We now give the basis for the two polarization 
four-vectors $\epsilon^1_j$ and $\epsilon^2_j$ of the photons
($j=1,2,3$, each outgoing photon has two polarizations available). 
These four-vectors satisfy
$\epsilon_j^1\cdot k_j=\epsilon_j^2 \cdot k_j=0$ 
(for each $j$ individually, no sum over $j$) and  are needed to analyze the
polarization-resolved cross sections. We take them as
\begin{align}
\label{polvecs}
\epsilon_j^1 &= 
(0,\vecb{\epsilon}_j^{\,1} )=
\left(0, \, \cos\theta_j  \cos\phi_j, \, 
\cos\theta_j \sin\phi_j, \, -\sin\theta_j\right) \,,
\nonumber
\\
\epsilon_j^2 &= 
(0,\vecb{\epsilon}_j^{\,2} )=
\left(0, \, -\sin\phi_j, \, \cos\phi_j, \, 0 \right) \,.
\end{align}
The superscript denotes either one of the two available polarizations.
%
%
\subsection{Matrix element and differential cross section}
The expression for the cross section of the three-photon Compton effect 
follows in a straightforward way from the usual 
Feynman rules of QED \cite{JaRo1980,ItZu2006}. 
The expression for the invariant matrix element $M_{\rm TC}$ reads
(TC stands for triple Compton)
\begin{equation}
\label{matrixelement}
M_{\rm TC} = \frac14 \, \frac{e^4 \, N_{\rm TC}}{(2\pi)^5} \,
\frac{1}{m^2\sqrt{E_i E_f \omega_0 \omega_1 \omega_2 \omega_3}} \,,
\end{equation}
with
\begin{align}
\label{defN}
N_{\rm TC}={}m^3
\sum_{\perm} 
&
u_{r_f}^\dagger(p_f)\gamma^0 
\hat{\epsilon}_{\perm(3)}\frac{\hat{q}_3 (\perm)+m}{q_3^2(\perm)-m^2}
\hat{\epsilon}_{\perm(2)}\frac{\hat{q}_2(\perm)+m}{q_2^2(\perm)-m^2}
\nonumber
\\&
\times
\hat{\epsilon}_{\perm(1)}\frac{\hat{q}_1(\perm)+m}{q_1^2(\perm)-m^2}
\hat{\epsilon}_{\perm(0)}u_{r_i}(p_i).
\end{align}
In Eq.~\eqref{defN}, the sum runs over all the $4!=24$ 
available permutations $\perm$ of
$(0,1,2,3)$ and describes the bosonic symmetrization
of the final state. One might think that, 
because of the presence of three indistinguishable 
particles in the final state, an additional combinatorial factor 
should have to be taken into account.
Indeed, a factor $1/3!$ must be inserted 
if we seek to calculate the total cross section 
for the triple scattering process~(see~Sec.~\ref{subsec:totcrosssection} below),
roughly speaking, because the ``wide-angle detectors''
needed for the theoretical calculation of the 
total cross section (with overlapping 
acceptor solid angles)
would otherwise detect the same photon more than once.
There is no need to add such a factor in
the differential cross section.
We here include this consideration 
because it might be important
for experiments in the future. 

\begin{figure}
\includegraphics[width=0.45\textwidth]{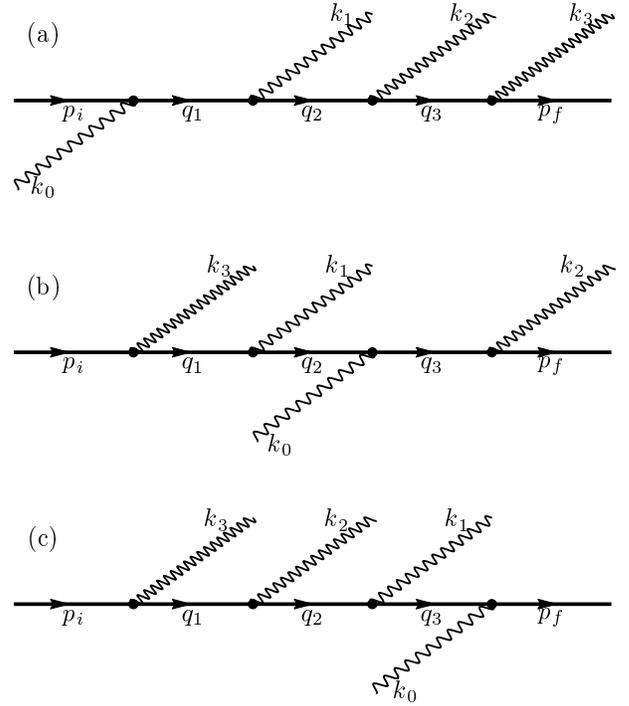}
\caption{\label{fig1}
Three Feynman diagrams out of the total of $4!=24$ which
contribute to the three-photon Compton effect. Diagram (a) corresponds to 
the permutation $\perm=(0,1,2,3)$, diagram~(b) to 
the permutation $\perm=(3,1,0,2)$, 
and diagram (c) to the permutation $\perm=(3,2,1,0)$.}
\end{figure}

Let us also give an example for the permutations 
entering Eq.~\eqref{defN}: E.g., if $\perm=(2,3,1,0)$, then $\perm(0)=2$,
$\perm(1)=3$, $\perm(2)=1$, $\perm(3)=0$, and so on. 
The momenta $q_n$ entering the propagators are calculated 
according to the equation
\begin{equation}
q_n(\perm)=p_i+\sum_{j=0}^{n-1}(-1)^{1-\delta_{0\perm(j)}}k_{\perm(j)} \,,
\end{equation}
which describes the momentum flow through the diagram
($\delta_{ij}$ is the Kronecker delta).
The zeroth photon with propagation four-vector $k_0$ 
adds to the momentum flow, while the three emissions
with $k_j$ ($j=1,2,3$) need to be subtracted. 
Each one  of the terms in Eq.~\eqref{defN} corresponds to one Feynman diagram,
three of which are exemplified in Fig.~\ref{fig1}.

According to the Feynman rules of QED,
the differential cross section follows from the 
matrix element~\eqref{matrixelement} as
\begin{align}
\label{diffcrosssection1}
& \frac{\d\sigma}{\d\omega_1 \d\omega_2 \d\omega_3 
\d^3 p_f \d\Omega_1 \d\Omega_2 \d\Omega_3} =
\\ 
&
(2\pi)^2 \frac{E_i \, \omega_0}{p_i \cdot k_0} \,
|M_{\rm TC}|^2 \, \omega_1^2 \, \omega_2^2 \, \omega_3^2 \,
\delta^{(4)}(p_i+k_0-\sum_{j=1}^3 k_j-p_f),
\nonumber
\end{align}
where $\d\Omega_j=\d\phi_j \d\theta_j\sin\theta_j$ is the 
infinitesimal solid angle of photon $j$.
In \eqref{diffcrosssection1}, both the integrations over 
$\d^3p_f$ and $\d\omega_3$ can be taken with the 
aid of the delta function, to yield 
\begin{subequations}
\begin{align}
\label{omega3}
\vecb{p}_f=& \; \vecb{p}_i+\vecb{k}_0-\sum_{j=1}^3\vecb{k}_j \,,
\\
\label{omega3b}
E_f=& \; E_i+\omega_0-\sum_{j=1}^3\omega_j \,,
\\
\label{omega3c}
\omega_3=& \;
\frac{p_i\cdot ( k_1+ k_2-k_0)+k_0\cdot( k_1+ k_2)-k_1\cdot k_2}%
{n_3\cdot (k_1+k_2-k_0-p_i)}.
\end{align}
\end{subequations}
For fixed values of the angles $\theta_j$, $\phi_j$,  
the condition $\omega_3 > 0$ defines the area of 
the  $\omega_1\omega_2$ plane in which 
the differential cross section is nonvanishing;
otherwise it is zero due to kinematic constraints.
The integration over $\d\omega_3$ introduces an additional factor
\begin{equation}
\frac{\d(E_f+\omega_3)}{\d\omega_3} = 1 +
\frac{\vecb{n}_3\cdot (\vecb{k}_1+\vecb{k}_2-\vecb{k}_0-\vecb{p}_i)+\omega_3}{E_f}.
\end{equation}
The final expression for the differential cross section of the three-photon Compton
effect, differential in the 6 angles and 2 energies of the emitted photons,
 still dependent on the one incoming and three outgoing photon
polarizations, and the electron spins, reads (in natural units)
\begin{align}
\label{diffcrosssectionfinal} 
& \frac{\d\sigma_{\rm TC}}%
{\d\omega_1 \, \d\omega_2 \, \d\Omega_1 \, \d\Omega_2 \, \d\Omega_3} =
\frac{\alpha^4}{(2\pi)^4} 
\frac{1}{m^4}
\frac{\omega_1\omega_2\omega_3}{ E_f \, (p_i \cdot k_0)} 
\nonumber\\[0.7ex]
& \quad \times
\left|\left(\frac{\d(E_f+\omega_3)}{\d\omega_3}\right)^{-1}\right|
|N_{\rm TC}|^2 \, \Theta(\omega_3) \, \Theta(E_f-m).
\end{align}
In
Eq.~\eqref{diffcrosssectionfinal}, $\omega_3$ should be replaced according to
Eq.~\eqref{omega3c}, and $p_f$ is to be replaced according to
$p_f=p_i+k_0-\sum_{j=1}^3 k_j$. 
The step
functions $\Theta(\cdot)$ at the end of Eq.~\eqref{diffcrosssectionfinal} 
are needed since there are values for the
angles that result in $\omega_3>0$ from Eq.~\eqref{omega3c}, but $E_f<m$. 

The cross section \eqref{diffcrosssectionfinal} diverges whenever either
$\omega_1$, $\omega_2$ or $\omega_3$ goes to zero. This is the well-known
infrared catastrophe of QED. In the current case, the divergences would cancel
against fourth-order (in $\alpha$) radiative corrections to the single and
double Compton effect.  The radiated energy, which is proportional to $\int \d
\omega_1 \int \d \omega_2  \int \d \omega_3 \, \omega_1 \omega_2  \omega_3
\d\sigma_{\rm
TC}/(\d\Omega_1\d\Omega_2\d\Omega_3\d\omega_1\d\omega_2\d\omega_3)$ is still
finite when integrated in the infrared.  While certainly an interesting subject
of study~\cite{BrFe1952,Su1956,YeFrSu1961,Mo1971}, such corrections will not be
considered in the present~\paper.  In general, radiative corrections to the
cross section are expected to be of order $\alpha$, or at the few-percent
level, since we consider photon energies (in the rest frame of the electron) of
at most $\omega_0=100$ MeV in this \paper{} (see
Sec.~\ref{subsec:totcrosssection} on the total cross section).  At high energy
$\omega_0\gg m$, infrared radiative corrections can be shown to be the dominant
ones \cite{YeFrSu1961}. The bremsstrahlung corrections in the 
exit channel, which are cancelled by the radiative corrections
in the infrared, are in our case of the order $c_{\rm
IR}=(\alpha/\pi)\ln(2\omega_0/m)\ln(2\omega_0m/\Delta^2)$, where $\Delta$ is
the energy resolution of the  detector. The precise value therefore depends on
the experimental setup~\cite{YeFrSu1961}.  Assuming that
$\omega_0/\Delta=100$, then $c_{\rm IR}\approx 0.06$ for $\omega_0=100$ MeV. In
all our examples for the differential cross section in
Secs.~\ref{subec:diffcrosssec} and \ref{subsec:diffcrosssecGeV}, however, the
energy scale is much smaller. In the electron rest frame, we have
$\omega_0/m\approx 0.4$ in Sec.~\ref{subec:diffcrosssec} and $\omega_0/m\approx
1$ in Sec.~\ref{subsec:diffcrosssecGeV}. In this case we expect $c_{\rm IR}=
(\alpha/\pi)\ln(2m^2/\Delta^2)\approx 0.02$.
So, the radiative corrections are expected not to exceed the level of a few percent.

Let us dwell on this point a little longer,
assuming the latter situation, where
$c_{\rm IR} \sim (\alpha/\pi)\ln(m/\Delta)$ 
up to multiplicative factors.
The theorem of Yennie, Frautschi and Suura~\cite{YeFrSu1961}
as well as the considerations of Sudakov~\cite{Su1956} imply that,
if the calculation were carried through to infinite loop 
order, the infrared divergences exponentiate according to
\begin{align}
& \sum_{n=0}^\infty \frac{(-1)^n}{n!} \,
\left[ \frac{\alpha}{\pi} 
\ln\left(\frac{m}{\Delta}\right)  \right]^n
= \exp \left[ -\frac{\alpha}{\pi} \ln\left(\frac{m}{\Delta} \right) \right],
\nonumber\\[0.33ex]
%
\end{align}
but, if the calculation is carried out only to a finite 
loop order, then the next higher-order terms 
(in $\alpha$) will yield radiative and bremsstrahlung corrections
on the percent level.

In an experiment, the detectors are always set up such as to
detect photons above a certain infrared threshold energy. 
For example, the experiment in~Ref.~\cite{MGBr1968} detected photons with an 
energy greater than 13\,keV.
Theoretically, we do the same thing, i.e., when integrating over the energy, 
we only include photon energies larger than a fixed energy threshold which
we label $\varepsilon$. 
Let us put $\omega_3 = \varepsilon$ in Eq.~\eqref{omega3c}.
We can then calculate the maximum energy 
$\omega_1^{\textrm{max}}$ of $\omega_1$ as a function 
of all the other variables using $k_1 = \omega_1 \, n_1$
and solving for $\omega_1$. Let us investigate fixed emission
angles and photon energies $\omega_\ell$, where $\ell=2$ if $j=1$, and $\ell=1$ if
$j=2$ (formally, $\ell = 3-j$). Then, one obtains for the 
maximimum energy $\omega_j^{\textrm{max}}$ of the $j$th photon
the expression
\begin{equation}
\label{omegaj_max}
\omega_j^{\textrm{max}} =
\frac{\varepsilon n_3 \cdot (k_\ell-p_i-k_0)+p_i\cdot k_0-k_\ell \cdot(p_i+ k_0)}
{n_j\cdot(p_i+k_0-k_\ell-\varepsilon n_3)}.
\end{equation}
The differential cross section integrated over the final photon energies will
depend on the infrared cutoff 
$\varepsilon$. Furthermore, since the specification of an energy
threshold depends on the observer frame, total cross sections are no longer
Lorentz invariant,  but the 
applicable threshold is fixed by the properties of the detectors used.

%
%
\subsection{Comparison to the double Compton effect}
\label{DCdisc}

In order to compare the cross section for the three-photon Compton effect with that
of the double Compton effect, we now take a step back,
and first give the expressions for the double (two-photon)  Compton
differential cross section. Although there exists an analytic expression
for the polarization- and spin-summed cross section~\cite{MaSk1952},
there is no analytic expression available for the polarization-resolved cross section.
The matrix element for the double Compton (DC) effect reads
\begin{equation}
\label{matrixelementDC}
M_{\rm DC}={}
e^3\frac{1}{(2\pi)^{\frac 7 2}}\frac{1}{m\sqrt{8E_i E_f \omega_0 \omega_1 \omega_2}}
N_{\rm DC},
\end{equation}
with
\begin{align}
\label{defNDC}
N_{\rm DC}={}m^2
\sum_{\permdc}
& 
u_{r_f}(p_f)\gamma^0 
\hat{\epsilon}_{\permdc(2)}\frac{\hat{q}_2 (\permdc)+m}{q_2^2(\permdc)-m^2}
\nonumber
\\&
\times
\hat{\epsilon}_{\permdc(1)}\frac{\hat{q}_1(\permdc)+m}{q_1^2(\permdc)-m^2}
\hat{\epsilon}_{\permdc(0)}u_{r_i}(p_i),
\end{align}
where  the sum runs over all the $3!=6$ permutations $\permdc$ of $(0,1,2)$. The 
momenta $q_n$ entering the propagators are defined similarly to the three-photon 
Compton case as 
\begin{equation}
q_n(\permdc)=p_i+\sum_{j=0}^{n-1}(-1)^{1-\delta_{0\permdc(j)}}k_{\permdc(j)},
\end{equation}
i.e.~the zeroth photon momentum flows in, the others flow 
out. The cross section, differential in $\omega_1$, $\Omega_1$ and $\Omega_2$,
follows  in analogy with Eq.~\eqref{diffcrosssectionfinal} above as 
\begin{align}
\label{sigmaDC}
\frac{\d\sigma_{\rm DC}}{\d\omega_1 \d\Omega_1\d\Omega_2}={}&
\frac{\alpha^3}{(2\pi)^2} 
\frac{1}{m^2}
\frac{\omega_1\omega_2}{E_f \, (p_i\cdot k_0)}
\left|\left(\frac{\d(E_f+\omega_2)}{\d\omega_2}\right)^{-1}\right|
\nonumber\\
&
\times |N_{\rm DC}|^2  \Theta(\omega_2)\Theta(E_f-m).
\end{align}
In \eqref{sigmaDC}, the final momentum of the electron is 
$p_f=p_i+k_0-k_1-k_2$, and 
\begin{equation}
\omega_2=\frac{p_i\cdot (k_1-k_0)  +k_0\cdot k_1}
{n_2\cdot(k_1-p_i-k_0)}.
\end{equation}
The Dirac-$\delta$ function generates a Jacobian factor of
$\d(E_f+\omega_2)/\d\omega_2 = 1 + 
[\omega_2+\vecb{n}_2\cdot(\vecb{k}_1-\vecb{p}_i-\vecb{k}_0)]/E_f$.
The double Compton cross section 
is used for comparison to the triple Compton 
effect in the following. The infrared cutoff $\varepsilon$ is 
used in complete analogy to the triple-Compton process.

%
%
\section{Numerical Examples}
\label{examples}
\subsection{Orientation}

We now  return to the three-photon Compton effect,
while using the discussion of the double (two-photon) Compton 
scattering process from Sec.~\ref{DCdisc} as a guide toward 
the comparison with lower-order processes.
We thus  numerically evaluate a number of examples of 
both the differential and the total cross section for experimentally 
realizable values of the parameters. The evaluation of the cross 
section is performed entirely numerically, assuming the 
representations given in Eqs.~\eqref{bispinordef} and \eqref{diracgammadef} 
for the gamma matrices and the electron spinors. Given the input parameters 
$\omega_0$, $E_i$, $\epsilon_0$, and $\epsilon_j$, $\theta_j$, $\phi_j$, $j=1,2,3$ 
(electron spin is always summed over), 
the evaluation of the matrix element is done by 
{\em explicit} matrix multiplication using the standard 
Dirac representation of the Clifford algebra given in Eq.~\eqref{diracgammadef}. 
This method is far preferable, because an analytic evaluation of the cross section by 
tracing out the Dirac-$\gamma$ matrices would result in an extremely long 
analytic expression which would not simplify (because we are
investigating the differential cross section)
and thus not be useful. The necessity to avoid an ``explosion'' in the 
number of terms in the intermediate expressions is particularly 
important, because we are
interested in polarization-resolved cross sections, in which case 
there are no simplifications at all in the analytic trace. On the
occasion, we also recall arguments given by us 
previously in Ref.~\cite{ScLoJeKe2007pra} 
which demonstrate that, for typical multiple scattering processes,
it is computationally faster 
to evaluate the matrix element by direct numerical matrix multiplication 
than to evaluate the analytic expressions that would otherwise result from 
the Dirac-$\gamma$ matrix trace.

One test of correctness of the numerical implementation of the cross section is
that of gauge invariance. The cross section \eqref{diffcrosssectionfinal} is
invariant under the gauge transformation 
\begin{equation}
\label{gaugetransformation}
\epsilon_j\to \epsilon_j + A \, k_j,
\end{equation}
for each $j=1,2,3$ separately.  Here, $A$ is an  arbitrary constant.

In the numerical integration of the differential cross section, the integration 
over $\omega_1$ and $\omega_2$ is done by a standard Romberg routine. 
By contrast, the integration over the emission angles of the photons is performed by 
Monte Carlo integration \cite{PrFlTeVe1993}. Monte Carlo integration is a well-established 
method for QED processes with a many-dimensional 
final state phase space~\cite{HuMuKe2010,DPHaKe2010}, pioneered 
by Mork \cite{Mo1967,JaMo1973}.  

\begin{figure}
\includegraphics[width=0.48\textwidth]{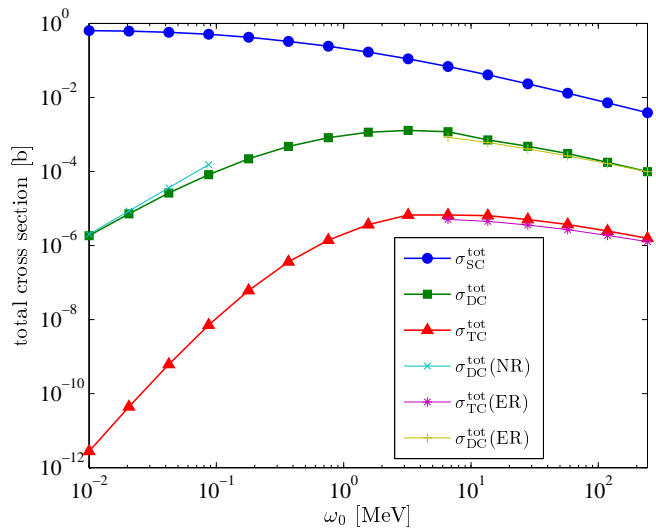}
\caption{\label{fig2}
\colol
The total cross section $\sigma^{\textrm{tot}}$ 
is plotted as a function of the initial photon
energy $\omega_0$, for  the single (SC), double (DC), and triple (TC) Compton
effect. The unit of barn is given 
as $1\, {\rm b}=10^{-24}\,{\rm cm}^2\approx 389.4^{-1}\,{\rm MeV}^{-2}$.
The non-relativistic (NR) DC approximation is from Ref.~\cite{Go1984}, and the
extreme relativistic (ER) approximations are taken from 
Ref.~\cite{MaMaDh1959}.  It
is assumed that $E_i=m$ and that the photon energy threshold is
$\varepsilon=\omega_0/50$.  Note the doubly logarithmic scale.}
\end{figure}

%
%
\subsection{Total cross section for $\bm{n}$-fold scattering}
\label{subsec:totcrosssection}

In order to get an impression of the order of magnitude of the total number of 
three-photon  events produced in an experiment, we first calculate
the total cross section
\begin{align}
\sigma_{\rm TC}^{\textrm{tot}} =& \;
\frac{ 1 }{3!4}
\sum_{\textrm{spin},\,\textrm{pol.}}
\int
\d\Omega_1\d\Omega_2\d\Omega_3
\nonumber
\\
&
\times 
\int\limits_{\omega_{1,2,3}>\varepsilon} \d\omega_1\d\omega_2
\frac{\d\sigma_{\rm TC}}{\d\omega_1 \d\omega_2  
\d\Omega_1 \d\Omega_2 \d\Omega_3} \,,
\end{align}
averaged over initial polarization and spin, and summed over final state
polarization and spin.  Here, in  contrast to the differential cross section,
the factor of $1/3!=1/6$ is inserted to compensate for the  double-counting of
equivalent angular configurations.  The electron is assumed to be initially at
rest, $E_i=m$, and we assume the threshold $\varepsilon=\omega_0/50$.  We aim
to compare to the total double Compton cross section, 
\begin{align}
\sigma_{\rm DC}^{\textrm{tot}}=
\frac{ 1 }{2!4}
\sum_{\textrm{spin},\,\textrm{pol.}}
\int
&
\d\Omega_1\d\Omega_2
\int\limits_{\omega_{1,2}>\varepsilon} \d\omega_1
\frac{\d\sigma_{\rm DC}}{\d\omega_1  \d\Omega_1 \d\Omega_2 },
\end{align}
and the total cross section $\sigma_{\rm SC}^{\textrm{tot}}$ of the usual, single 
Compton (SC) effect, which is known analytically as \cite{JaRo1980}
\begin{align}
 \sigma_{\rm SC}^{\textrm{tot}} = 2\pi\frac{\alpha^2}{m^2}
\Bigg\{ & \frac{1+\omega}{\omega^3}
\left[\frac{2 \omega (1+\omega)}{1+2 \omega} -\ln(1+2\omega)\right]
\nonumber
\\
& + \frac{\ln(1+2\omega)}{2\omega} -\frac{1+3\omega}{(1+2\omega)^2 }\Bigg\},
\end{align}
with $\omega=\omega_0/m$. 
As is well known, this cross section 
approaches a constant in the limit $\omega \to 0$, 
which reads as 
\begin{equation}
\sigma_{\rm SC}^{\textrm{tot}} = 
\frac{8 \pi \alpha^2}{3 m^2} \, 
\left[ 1 - 2 \, \omega + \calO\left(\omega^2\right) 
\right]  \,.
\end{equation}
The constant limit for $\omega_0 \to 0$ can be discerned in
Fig.~\ref{fig2}, where we  show the total cross sections for the single (SC),
double (DC), and triple (TC)  Compton effect. For the DC case, we have included
a comparison with the non-relativistic result from Ref.~\cite{Go1984},
\begin{equation}
\sigma_{\rm DC}^{\rm tot}({\rm NR})= C_{\rm DC}^{\rm NR}   \frac{\alpha^3}{m^2}
\left(\frac{\omega_0}{m}\right)^2.
\end{equation}
With our convention of
$\varepsilon=\omega_0/50$ for the photon energy threshold, the
constant $C_{\rm DC}^{\rm NR}\approx 9.1$.  For TC, a numerical fit of the
calculated points for $\omega_0<0.1$\,MeV gives $\sigma_{\rm TC}^{\rm
tot}\propto \omega_0^n$ with $n\approx 3.6$.  For low energies $\omega_0\ll m$,
$\sigma_{\rm TC}^{\rm tot}$ should be proportional to
$\omega_0^4/m^6$, like
\begin{equation}
\label{sigmaTCNR}
\sigma_{\rm TC}^{\rm tot}({\rm NR})=C_{\rm TC}^{\rm NR}\frac{\alpha^4}{m^2}
\left(\frac{\omega_0}{m}\right)^4 \,.
\end{equation}
By matching expression \eqref{sigmaTCNR} with $\sigma_{\rm TC}^{\rm tot}$
calculated at $\omega_0=10^{-2}$ MeV (the leftmost point in
Fig.~\ref{fig2}), we obtain $C_{\rm TC}^{\rm NR}\approx 4.5$ for the constant 
prefactor.  

In the
extreme relativistic (ER) limit, the total cross sections for the two-photon and
three-photon Compton effect have been calculated in~\cite{MaMaDh1959} in the
approximation $\omega_0\gg m$, $\omega_1\gg m$, $\omega_j\ll m$ for $j>1$, and
$\omega_0\gg\omega_1$. The result is~\cite{MaMaDh1959}
\begin{equation}
\label{sigmatotER}
\sigma^{\rm tot}({\rm ER}) =
\frac{1}{n!}\left[\frac{\alpha}{\pi} \ln \left(\frac{2\omega_0}{m}\right)
\ln\left(\frac{\varepsilon_{\rm up}}{\varepsilon_{\rm low}} \right) \right]^n \,
\sigma_{\rm SC}^{\rm tot},
\end{equation}
where $n=1$ for DC and $n=2$ for TC, and $\varepsilon_{\rm low, up}$ are the
lower and upper limits for the energy of the soft photons $\omega_{j>1}$. The
result \eqref{sigmatotER} is interesting, since it implies that at extremely  high
energies, the total cross sections of DC and TC can exceed that of SC. However, 
the energy scale at which this occurs is so high
(the energy scale is of the order of the Landau pole in QED), 
so that this question is rather academic.  Although
the assumptions leading to the formula \eqref{sigmatotER} do not hold in our
case, since we assume that the photon energy threshold varies with the incoming
photon energy as $\varepsilon=\omega_0/50$, and the soft-photon requirement
$\omega_{j>1}\ll m$ becomes impossible to satisfy at high energies, we have
still included the total cross section obtained from Eq.~\eqref{sigmatotER} in
Fig.~\ref{fig2}.  In calculating $\sigma_{\rm DC, TC}^{\rm tot}({\rm ER})$,
we assumed that $\varepsilon_{\rm up}/\varepsilon_{\rm low}=5$. We can see 
from Fig.~\ref{fig2} that the expression \eqref{sigmatotER} well approximates 
the calculated $\sigma^{\rm tot}_{\rm DC,TC}$ for $\omega_0\gtrsim10$~MeV.
 The decrease in the cross section with increasing 
$\omega_0$ in the fully relativistic regime is due to the fact that 
%
%
%
\begin{equation}
\label{sigmaTCER}
\sigma_{\rm SC}^{\rm tot}\approx\frac{\pi\alpha^2}{m}
\frac{1}{\omega_0} 
\ln \left(\frac{2\omega_0}{m}\right)
\end{equation}
for large $\omega_0\gg m$.
Numerically, our  values for $\sigma_{\rm DC}^{\rm tot}$ disagree
with those calculated in \cite{RaWa1971}, but one has to 
be aware that in Ref.~\cite{RaWa1971},  a different convention  for
the photon energy threshold is employed.

From Fig.~\ref{fig2}, we can infer a few interesting facts. Contrary to
$\sigma_{\rm SC}^{\rm tot}$, which monotonically decreases with increasing
$\omega_0$,  $\sigma_{\rm DC}^{\rm tot}$ and $\sigma_{\rm TC}^{\rm tot}$ peak
at a certain value of $\omega_0=\omega_0^{\rm max}$. The data in
Fig.~\ref{fig2} roughly 
give the same value of $\omega_0^{\rm max}= 3.2$\,MeV for
both DC and TC, with $\sigma_{\rm DC}^{\rm tot}(\omega_0^{\rm max})=1\times
10^{-3}$\,b and $\sigma_{\rm TC}^{\rm tot}(\omega_0^{\rm max})=7\times 10^{-6}$\,b.

\begin{figure}
\includegraphics[width=0.5\textwidth]{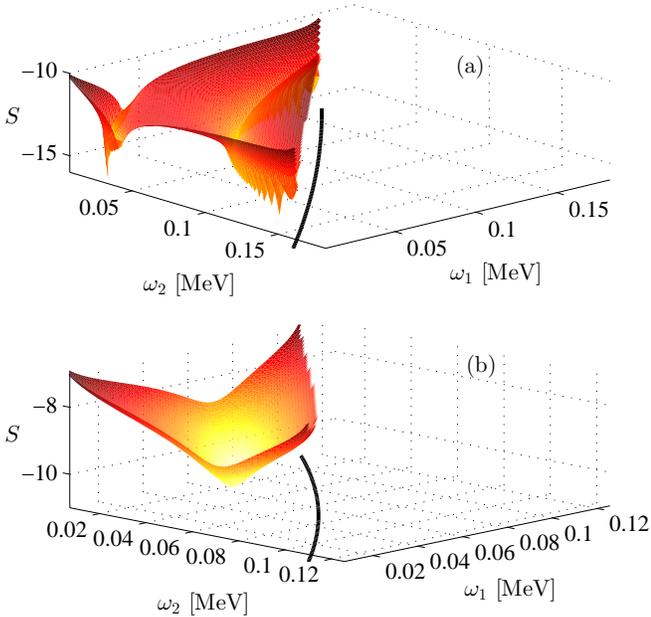}
\caption{\label{fig3} 
\colol Differential cross section as a function 
of  $\omega_1$ and $\omega_2$. We have $\omega_0=180$~keV, $E_i=m$, 
$\varepsilon=\omega_0/50$, and $\phi_j=2j\pi/3$, $j=1,2,3$. The value of the polar angle is 
 $\theta_{1,2,3}=1/2$ in panel (a) and $\theta_{1,2,3}=2$ in panel (b),
which corresponds to a triple scattering events in the forward and 
backward cones, respectively (relative to the incoming photon).
The actual quantity plotted is $S$ (the decadic logarithm of the differential cross section), 
as defined in Eq.~\eqref{defS}.
For the polarizations of the final photons, 
we have $(\vecb{\epsilon}_1,\vecb{\epsilon}_2,\vecb{\epsilon}_3)=
(\vecb{\epsilon}_1^{\,1}, \vecb{\epsilon}_2^{\,1},
\vecb{\epsilon}_3^{\,1})$ in both panels, and the incoming photon is polarized in the 
$x$-direction.
The thick, black line shows the curve along which the energy of 
photon three is at the assumed detector threshold, 
$\omega_3=\varepsilon=3.6$~keV. This curve can be calculated 
according to Eq.~\eqref{omegaj_max}.  }
\end{figure}

\begin{figure*}
\includegraphics[width=0.99\textwidth]{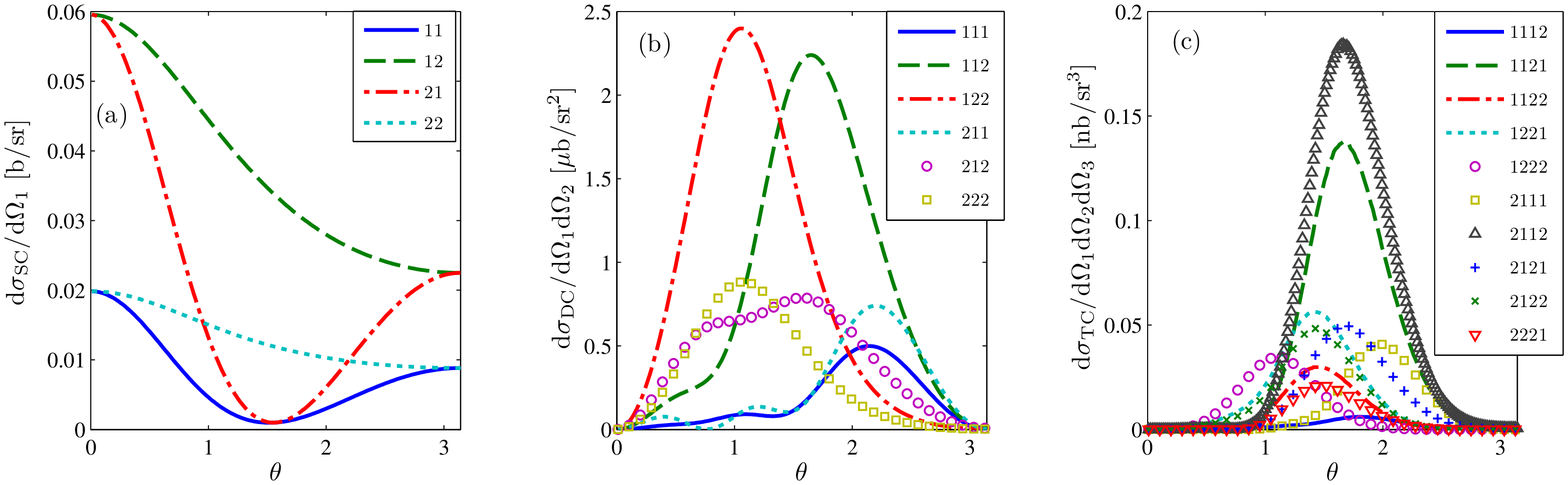}
\caption{\label{fig4} 
\colol 
Comparison of the (a) single, (b) double, and (c) triple   Compton 
differential cross sections at $\omega_0=180$~keV, $E_i=m$, 
integrated over photon energies larger than $\varepsilon=\omega_0/50$. It is assumed 
that $\theta_j=\theta$, $\phi_j=2j\pi/3$, with $j=1$ for SC, $j=1,2$ for DC, and $j=1,2,3$ for TC. 
The indices $ij$, $ijk$, and $ijk\ell$ in the legends indicate the polarizations of the photons 
as $(\vecb{\epsilon}_0,\vecb{\epsilon}_1,\vecb{\epsilon}_2,\vecb{\epsilon}_3)=
(\vecb{\epsilon}_0^{\,i}, \vecb{\epsilon}_1^{\,j},
\vecb{\epsilon}_2^{\,k},\vecb{\epsilon}_3^{\,\ell})$ for TC, 
and correspondingly for DC and SC
(with $i,j,k,\ell=1,2$).
For symmetry reasons, in DC [panel (b)], $121$ 
has the same curve as $112$, and $221$ has the same 
curve as $212$. Similarly, in TC [panel (c)], $1111$ has the same curve as $2111$, $1212=1122$, 
$1211=1121$, $1222=2222$, $2211=2121$, and $2212=2122$.  
}
\end{figure*}

%
%
\subsection{180 keV photons on stationary electrons}
\label{subec:diffcrosssec}

Our first example for the differential cross section is taken at $\omega_0=180$
keV, $\varepsilon=\omega_0/50$, and $E_i=m$. This situation seems favorable for
an experimental verification of the three-photon Compton effect which goes
beyond that in Ref.~\cite{MGBr1968}, since photons of this energy are available
at synchrotron radiation sources \cite{SaEtAl2001} with a high photon flux, and
stationary targets allow for a high electron density.  The total cross section
is calculated to be $\sigma_{\rm TC}^{\rm tot}=6\times 10^{-8}$\,b, which is
rather low, but can be compensated for by the aforementioned high photon flux
and large number of target electrons.  If we  assume  a photon flux of $2\times
10^{12}/\textrm{s}$ (see Ref.~\cite{SaEtAl2001}), and a 0.1~mm thick Al foil as
the target, then we obtain about 900 photon triplets per second.

In Fig.~\ref{fig3}, we show the differential cross section as a function of
$\omega_1$ and $\omega_2$, for fixed emission angles of the photons and a
particular set of final polarization vectors.  For plotting purposes, we define
the dimensionless quantity $S$ as 
\begin{equation}
\label{defS}
S=\log_{10}\left(\frac{1}{2} \sum_{\rm spin}\frac{\d \sigma_{\rm TC}}
{\d\Omega_1 \d\Omega_2\d\Omega_3 \d\omega_1\d\omega_2}
\frac{{\rm MeV}^{2}\, {\rm sr}^3}{{\rm b} }\right),
\end{equation}
i.e., the decadic logarithm of the differential cross section 
averaged over the incoming, and summed over the outgoing
electron spin, in units of ${\rm b}\,{\rm MeV}^{-2}\,{\rm sr}^{-3}$. For later
use we also define the corresponding polarization-summed quantity
\begin{equation}\label{defSbar}
\overline{S}=\log_{10}\left(\frac{1}{2} \sum_{\rm spin,\;pol.}\frac{\d \sigma_{\rm TC}}
{\d\Omega_1 \d\Omega_2\d\Omega_3 \d\omega_1\d\omega_2}
\frac{{\rm MeV}^{2}\, {\rm sr}^3}{{\rm b} }\right) \,,
\end{equation}
 where both electron spins and photon 
polarizations are summed over.
In Fig.~\ref{fig3}, the azimuthal angles of the three detectors are assumed to
describe a ``Mercedes-star'' configuration with $\phi_j=2j\pi/3$ for $j=1,2,3$.
The spin of the electron is summed over.   
For those values of $\omega_1$ and
$\omega_2$ which would otherwise give rise to $\omega_3 < \varepsilon$ according
to Eq.~\eqref{omega3c}, we have set the differential cross section to zero. The
line at which $\omega_3=\varepsilon$ (indicated with a thick, black line in
Fig.~\ref{fig3}) can be calculated with the help of Eq.~\eqref{omegaj_max}.  In
the current case, we have $n_1\cdot(p_i+k_0-\varepsilon n_3)\gg n_1\cdot k_2$,
which implies that the denominator in Eq.~\eqref{omegaj_max} is almost
constant, and consequently $\omega_1^{\textrm{max}}(\omega_2)$ becomes an
almost linear function of $\omega_2$.

Measuring  the 5-fold differential cross section of  the three-photon Compton effect 
would require fixing three photon detectors at different positions in 
space and in addition applying a spectrometer.
We can see that the patterns in the $\omega_1$$\omega_2$ 
plane and the overall magnitude of the differential 
cross section are different by several orders of 
magnitude depending on whether 
the final photons are emitted in a typical region within 
the forward cone
[$\theta=1/2$,  Fig.~\ref{fig3}(a)] or 
in the backward cone relative to the 
incoming photon [$\theta=2$,  Fig.~\ref{fig3}(b)].

In Fig.~\ref{fig4}, 
we present the differential cross section integrated over energy,
\begin{align}
\frac{\d\sigma_{\rm TC}}{\d\Omega_1 \d\Omega_2 \d\Omega_3}=& \;
\frac{1}{2}\sum_{\rm spin} \;\; \int\limits_{\omega_1,\omega_2,\omega_3>\varepsilon} 
\d\omega_1 \d\omega_2 
\nonumber
\\
&
\times
\frac{\d \sigma_{\rm TC}}{\d\Omega_1 \d\Omega_2 \d\Omega_3\d \omega_1\d \omega_2},
\end{align}
and similarly for DC. The polarization-resolved differential cross section for SC 
is known analytically. For $E_i=m$, we have~\cite{JaRo1980}
\begin{equation}
\frac{\d \sigma_{\rm SC}}{\d\Omega_1}=\frac{1}{4}\left(\frac{\alpha}{m}\right)^2 
\left(\frac{\omega_1}{\omega_0}\right)^2
\left(\frac{\omega_1}{\omega_0}+\frac{\omega_0}{\omega_1}-2
+4(\vecb{\epsilon}_1\cdot \vecb{\epsilon}_0)\right),
\end{equation}
where $\omega_1=\omega_0/[1+(\omega_0/m)(1-\cos\theta_1)]$.

\begin{figure}
\includegraphics[width=0.5\textwidth]{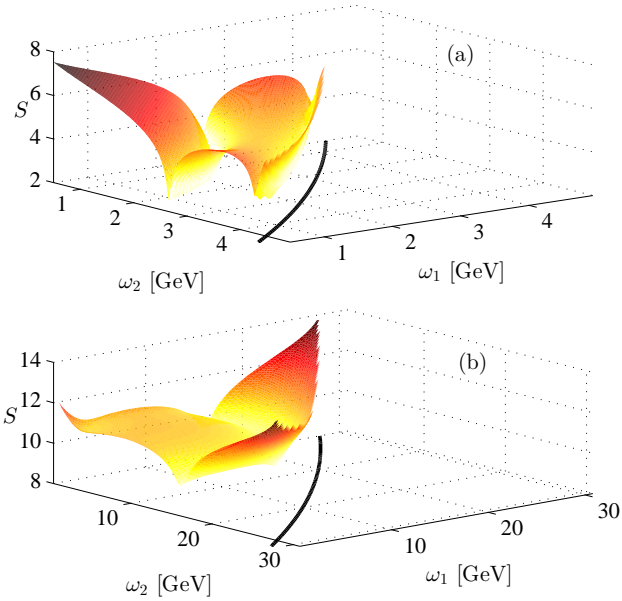}
\caption{\label{fig5} 
\colol 
Differential cross section as a function 
of  $\omega_1$ and $\omega_2$. We have $\omega_0=2.5$~eV, $E_i=50$~GeV, 
$\varepsilon=E_i/100$, and $\phi_j=2j\pi/3$ (again,
a ``Mercedes-star'' configuration of the detectors), for $j=1,2,3$. 
In panel (a), we have $\theta_{1,2,3}=\pi-4\times 10^{-5}$, and in (b),  
$\theta_{1,2,3}=\pi-7\times 10^{-6}$.
For visualization purposes, the decadic logarithm $S$ [defined in Eq.~\eqref{defS}] 
of the differential cross section is shown.
The polarizations of the final photons are given as
$(\vecb{\epsilon}_1,\vecb{\epsilon}_2,\vecb{\epsilon}_3)=
(\vecb{\epsilon}_1^{\,2},\vecb{\epsilon}_2^{\,2},\vecb{\epsilon}_3^{\,1})$,
while the incoming photon is polarized in the $x$-direction.
The black, thick line corresponds to the curve along which the energy of 
photon three is at the assumed detector threshold, 
$\omega_3=\varepsilon=500$~MeV.
}
\end{figure}

\begin{figure*}
\includegraphics[width=0.99\textwidth]{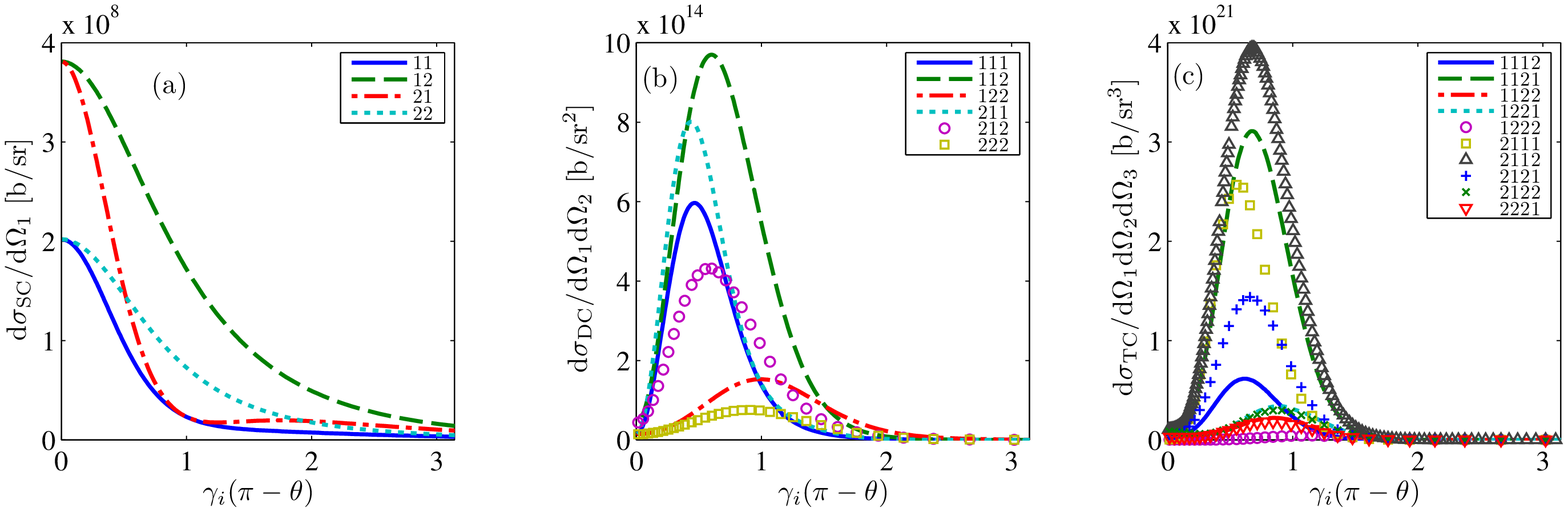}
\caption{\label{fig6} \colol Comparison of the (a) single, (b) double, and (c)
triple   Compton differential cross sections at $\omega_0=2.5$~eV, $E_i=50$~GeV
(backscattering from an incoming electron) in the laboratory frame, integrated
over photon energies larger than $\varepsilon=E_i/100$. It is assumed that
$\theta_j=\theta$, $\phi_j=2j\pi/3$, with $j=1$ for the SC, $j=1,2$ for the DC, and
$j=1,2,3$ for the TC effect.  The differential cross sections are plotted against
$\gamma_i(\pi-\theta)$, where $\gamma_i=E_i/m\approx 9.8\times 10^4$, and
$\gamma_i(\pi-\theta)=0$ implies that the photons are completely backscattered,
i.e., emitted anti-parallel to the incoming photon.  The indices $ij$, $ijk$,
and $ijk\ell$ in the legends indicate the polarizations of the photons as
$(\vecb{\epsilon}_0,\vecb{\epsilon}_1,\vecb{\epsilon}_2,\vecb{\epsilon}_3)=
(\vecb{\epsilon}_0^{\,i}, \vecb{\epsilon}_1^{\,j},
\vecb{\epsilon}_2^{\,k},\vecb{\epsilon}_3^{\,\ell})$ for TC, and
correspondingly for DC and SC (with $i,j,k,\ell=1,2$).  For symmetry reasons,
in DC [panel (b)], $121$ has the same curve as $112$, and $221$ has the same
curve as $212$. Similarly, in TC [panel (c)], $1111$ has the same curve as
$2111$, $1212=1122$, $1211=1121$, $1222=2222$, $2211=2121$, and $2212=2122$.
Note also that in (c), the $2221$ curve almost overlaps the $1122$ curve, and
the $2122$ curve almost overlaps the $1221$ curve.  }
\end{figure*}

%
%
\subsection{Laser photons on GeV electrons}
\label{subsec:diffcrosssecGeV}

In this example, we exploit the kinematics of a Compton backscattering setup,
which would allow for the creation of entangled photon triplets in the GeV
range.  The incoming electron is no longer at rest.  We take an
ultrarelativistic electron beam with $E_i=50$ GeV, and a visible laser photon
beam with $\omega_0=2.5$ eV (corresponding to a laser wavelength of 496 nm).
These parameters are close to those of the well-known
experiment~\cite{BuEtAl1997} performed at SLAC more than 15 years ago.  Here,
we  have in mind laser pulses which are not intense enough to induce
relativistic multi-photon processes, so that the scattering of a single laser
photon gives the dominant contribution to the cross section. This limits the
laser intensity to $I\lesssim 10^{17}$~W/cm$^2$.  In terms of the classical
nonlinear parameter $\xi=|e|F_{\rm peak}/(\omega_0 m)$, where $F_{\rm peak}$ is
the peak electric field of the laser~\cite{DPMuHaKe2012}, we have $\xi \approx
0.1$ for $I= 10^{17}$~W/cm$^2$ at a $\omega_0=2.5$~eV laser light.  We also note
that despite the high value of $E_i$, the incoming photon energy $\omega'_0$ in
the rest frame of the electron is $\omega'_0\approx 2\omega_0 E_i/m=0.5$~MeV,
which is below the $\e^+\,\e^-$ pair production threshold of $4m\approx 2$~MeV,
so that there is no  background connected with the creation of $\e^+\,\e^-$
pairs.  Strong-field (multi-photon) pair
production~\cite{HuMuKe2010,DPMuHaKe2012,Il2011} can also be neglected, since the
strong-field QED parameter $\chi=\xi k_0\cdot p_i/m^2\approx 0.1$ is much
smaller than unity. 

In this situation, due to the high gamma factor $\gamma_i=E_i/m$ of the electron beam, the 
photons are emitted in a narrow cone $\theta_j\sim \pi-1/\gamma_i$ around the axis 
of the incoming electron momentum $\vecb{p}_i$. In addition, the emitted photons 
can acquire high energy, the maximum energy for emission in the backward direction 
being given by the relativistic limit of the 
Compton formula as $4\omega_0\gamma_i^2$. 

The evaluation of the differential cross section becomes numerically
problematic due to the extreme parameter values, if the calculation is
performed in the laboratory  frame. It is instead advantageous to perform the
numerical calculation in the frame where the electron is initially at rest, and
then Lorentz transform the computed quantities into the laboratory frame.
Differential cross sections transform as
\begin{equation}
\frac{\d \sigma_{\rm TC}}{\d\Omega_1\d\Omega_2\d\Omega_3\d\omega_1\d\omega_2}
=J^{-1} \,
\frac{\d \sigma'_{\rm TC}}{\d\Omega'_1\d\Omega'_2\d\Omega'_3\d\omega'_1\d\omega'_2},
\end{equation}
with the relativistic Jacobian
\begin{equation}
J=\frac{(1-\beta_i^2)^2}{(1-\beta_i\cos\theta'_1) \,
(1-\beta_i\cos\theta'_2) \, (1-\beta_i\cos\theta'_3)^2},
\end{equation}
where we have denoted rest-frame quantities with a prime, 
and $\beta_i=\sqrt{1-1/\gamma_i^2}$.
For cross sections differential only in the angles we have instead
the Jacobian
\begin{equation}
\label{Jacobian_angles}
\tilde{J}=\frac{(1-\beta_i^2)^3}{(1-\beta_i\cos\theta'_1)^2 \,
(1-\beta_i\cos\theta'_2)^2 \, (1-\beta_i\cos\theta'_3)^2}.
\end{equation}
The photon energy threshold $\varepsilon$ is still 
fixed in the laboratory frame. Therefore, 
when doing the integration over $\omega'_1$ and $\omega'_2$, we set the cross section to 
zero if the lab-frame value of the photon energy is below the threshold, 
i.e., we impose the condition
\begin{equation}
\omega_j=\gamma_i(1-\beta_i\cos \theta'_j) \, \omega'_j>\varepsilon \,,
\end{equation}
for $j=1,2,3$.

Because the values for the total cross section shown in Fig.~\ref{fig2} only apply 
to an electron at rest, and with a different convention for the photon energy threshold, 
we have calculated anew the total cross section for the current example. We get
\begin{equation}
\label{fig2_lowEw0}
\sigma_{\rm TC}^{\rm tot}=6\times 10^{-7}\,{\rm b}
\end{equation}
for $\omega_0=2.5$~eV, $E_i=50$~GeV, and $\varepsilon=E_i/100=500$~MeV.
The value \eqref{fig2_lowEw0} coincides with the value of 
$\sigma_{\rm TC}^{\rm tot}(\omega'_0=0.5\,{\rm MeV},E_i=m)=6\times 10^{-7}$~b, which 
can be obtained from interpolation of the points in Fig.~\ref{fig2}.
Assuming an electron bunch containing $10^9$ electrons, 
 a laser intensity of $10^{17}$~W/cm$^2$, a pulse length 
of 100~fs, and perfect transverse overlap of the laser pulse and  the electron bunch, we obtain 
about 15 triple photon events per collision.

Examples of the fully differential cross section in the laboratory 
frame are shown in Fig.~\ref{fig5}. The differential cross sections integrated over 
the final photon energies are displayed in Fig.~\ref{fig6}. Due 
to the small value of the Jacobian \eqref{Jacobian_angles} at large $\gamma_i$, the 
DC and TC differential cross sections become very large in the lab frame.
However, numerically,
the total, integrated cross section for the triple scattering  
is not large [see Eq.~\eqref{fig2_lowEw0}].

%
%
\section{Multipartite entanglement}
\label{Sec:Entanglement}

The three photons in the final state of the triple Compton effect are emitted coherently, 
during the same coherence interval, and they are therefore quantum mechanically 
correlated, or entangled. Currently, a lot of effort is being invested
into the creation of 
controllable entangled quantum states of three or more particles in the 
laboratory~\cite{GrHoShZe1990,KeEtAl1998}. 
The conventional way of 
creating double or triple states of entangled photons is by nonlinear 
down-conversion in a crystal~\cite{PaEtAl2000,EiEtAl2003,%
GrBo2008,WeOhDu2010,%
AnGeKr2011,CoGPUr2011,%
DoBoBoBeLe2012}, 
while it is only recently 
that direct production of a triple photon state was successful~\cite{HuEtAl2010}.
With the current study, we propose the three-photon Compton effect as an alternative source of 
entangled triplets of photons. No nonlinear medium is required. 
Whether or not  the three-photon Compton effect is going to be effective  as a source of entangled photons depends on the 
experimental setup and the optimization thereof. Here, we limit ourselves to 
pointing out that the emitted three photons are entangled, 
and to an investigation of the degree of entanglement.

In principle, the emitted photons are entangled in 
all of their physical degrees of freedom: energies, angles, and 
polarization. The case which has 
been mostly investigated in other areas so far is that 
of entangled qubits, i.e., of entangled states within a well-defined 
manifold of discrete states (such as spin or polarization states).
Here, therefore, we study the polarization entanglement of the photons. 
The starting point is the density matrix $\rho$, which has elements    
\begin{equation}
\langle \lambda_1 \lambda_2 \lambda_3 | \rho | \lambda'_1 \lambda'_2 \lambda'_3 \rangle =
\kappa\sum_{\rm spin}  M_{\rm TC}(\lambda_1\lambda_2\lambda_3) 
M_{\rm TC}^\ast( \lambda'_1 \lambda'_2 \lambda'_3).
\end{equation}
The invariant matrix element for triple scattering  $M_{\rm TC}$ is given
in Eq.~\eqref{matrixelement}. The prefactor $\kappa$ is 
fixed by the normalization condition $\tr \rho =1$.
We have written 
\begin{equation}
M_{\rm TC}(\lambda_1\lambda_2\lambda_3)=
M_{\rm TC}(\vecb{\epsilon}_1=\vecb{\epsilon}_1^{\,\lambda_1},
\vecb{\epsilon}_2=\vecb{\epsilon}_2^{\,\lambda_2},
\vecb{\epsilon}_3=\vecb{\epsilon}_3^{\,\lambda_3}),
\end{equation} 
with $\lambda_j\in \{1,2\}$. 
When the state vectors $|\lambda_1\lambda_2\lambda_3\rangle$ are 
expressed as column vectors with $2^3=8$ entries, 
the density matrix $\rho$ becomes an $8\times 8$ 
matrix.
 
Given a density matrix $\rho$, it is a highly non-trivial problem to determine whether 
$\rho$ contains genuine multipartite entanglement or not 
\cite{SeUf2001,ToGuSeUf2005,BaGiLiPi2011,%
JuMoGu2011prl,%
JuMoGu2011pra,BrShVe2012,To2012,WuEtAl2012}. 
A $2\times 2\times 2$ system like the current one is 
considered to be genuinely multipartite entangled if its density matrix $\rho$ cannot 
be written in the form \cite{JuMoGu2011pra}
\begin{align}
\label{separable_rho}
\rho =& \;
p_1\sum_j q_1^j |\Lambda^j_1\rangle \langle \Lambda_1^j|   \otimes
|\Gamma_{23}^j\rangle \langle \Gamma_{23}^j |+
\nonumber
\\
&
p_2\sum_j q_2^j |\Lambda^j_2\rangle \langle \Lambda_2^j|   \otimes
|\Gamma_{13}^j\rangle \langle \Gamma_{13}^j |+
\nonumber
\\
&
p_3\sum_j q_3^j |\Lambda^j_3\rangle \langle \Lambda_3^j|   \otimes
|\Gamma_{12}^j\rangle \langle \Gamma_{12}^j |,
\end{align}
where $p_j$, $q_\ell^j$ are positive numbers satisfying 
$\sum_{j=1}^3 p_j =\sum_{j}  q_\ell^j=1$ for any $\ell=1,2,3$.
A state vector $|\Gamma^j_{k\ell}\rangle$ 
represents a general entangled state of photon $k$ and $\ell$, while a state 
$|\Lambda^j_n\rangle$ denotes a general one-photon state of photon $n$. 
Intuitively, a state $\rho$ is considered to be tripartite entangled if it cannot 
be written as  a sum of states that can be factorized   into states with less entanglement.
However, given a density matrix $\rho$, there is currently no efficient  algorithm
to decide whether or not $\rho$ can be written in the form~\eqref{separable_rho}.

In Refs.~\cite{JuMoGu2011prl,JuMoGu2011pra}, an algorithm was proposed, which
is able to detect ``almost all'' genuinely tripartite entangled states.  It
works as follows. Under the condition that the matrices $\opP_s$, $\opQ_s$,
$\mathbbm{1}-\opP_s$ and $\mathbbm{1}-\opQ_s$ do not have any negative
eigenvalues (which can be written in short form as $0\le \opP_s,\; \opQ_s \le
\mathbbm{1}$), we search for the maximum value of 
\begin{equation}
\label{entanglement_measure}
\tau(\rho)=-\tr (W \rho)
\end{equation} 
by varying $\opP_s$ and $\opQ_s$, where $W$ is assumed to be 
a function of $\opP_s$ and $\opQ_s$. 
Because $\mathbbm{1}-\opP_s$ and $\mathbbm{1}-\opQ_s$ have no negative eigenvalues,
in particular, one cannot take the entries of $\opP_s$ to be arbitarily 
large and positive, since then
$\mathbbm{1}-\opP_s$ would have negative eigenvalues.
In concrete terms, the entanglement witness $W$ 
in \eqref{entanglement_measure} is given as
\begin{equation}
\label{witnessW}
W=\opP_s+\opQ_s^{T_s}
\end{equation}
for all subsets $s\in\{1,2,3,12,13,23\}$, and 
$T_s$ denotes the partial 
transpose with respect to the subset $s$~\cite{LeKrCiHo2000}. If we write
\begin{equation}
\varrho=\sum_{ijk\ell mn=1}^2\varrho_{ijk\ell mn}
|i\rangle\langle j|\otimes
|k\rangle\langle \ell|\otimes
|m\rangle\langle n|
\end{equation}
for a generic density matrix, 
then, for example, the partial transpose with respect to $s=3$ is 
\begin{equation}
\varrho^{T_3}=\sum_{ijk\ell mn=1}^2\varrho_{ijk\ell mn}
|i\rangle\langle j|\otimes
|k\rangle\langle \ell|\otimes
|n\rangle\langle m|,
\end{equation}
and similarly for other values of $s$.  In general, an entanglement witness $W$
is a Hermitian matrix such that the trace $\tr(W\rho)$ is negative for at least
one entangled state $\rho$, and positive for all non-entangled states.  The
normalization of $W$ is limited by the positive eigenvalue condition $0\le
\opP_s,\; \opQ_s \le \mathbbm{1}$. If one is only interested in detecting
whether $\rho$ is entangled or not, and no quantitative measure of entanglement
is needed, the condition $\opP_s,\; \opQ_s \le \mathbbm{1}$ is replaced with
the normalization condition $\tr W=1$ \cite{JuMoGu2011prl,JuMoGu2011pra}. If
the maximum of $-\tr(W\rho)$ with $W=\opP_s+\opQ_s^{T_s}$ for all $s$ is
positive, the state $\rho$ is genuinely entangled.  It was shown in
\cite{JuMoGu2011prl,JuMoGu2011pra} that an entanglement witness on the form
\eqref{witnessW} can be used to detect a large class of genuinely entangled
states which are not so-called PPT mixtures. However, there are some genuinely
entangled states that are not detected by the algorithm.  The form
\eqref{witnessW} moreover permits the optimization of
\eqref{entanglement_measure} to be solved by the methods of convex optimization
theory, for which there exist standard software packages \cite{St1999}.  We
refer to~\cite{JuMoGu2011prl,JuMoGu2011pra} for further details about the
algorithm. 

The value of $\tau(\rho)$  is a measure of the tripartite entanglement present
in $\rho$. If  $\rho$ can be written in the form \eqref{separable_rho}, then
$\tau(\rho)=0$, and the state is not genuinely entangled. The reverse is not
true in general, i.e., even if $\tau(\rho)=0$, the state $\rho$ could still be
genuinely entangled.  The maximum value  of $\tau$ can be shown to be $1/2$
\cite{JuMoGu2011pra}. An example of a state which has $\tau(\rho)=1/2$ is the
Greenberger-Horne-Zeilinger (GHZ) state
$\rho_{\textrm{GHZ}}=|\textrm{GHZ}\rangle \langle \textrm{GHZ}|$, with
$|\textrm{GHZ}\rangle =(|111\rangle+|222\rangle)/\sqrt{2}$ (see
Ref.~\cite{GrHoShZe1990}), and the same (maximum) value of $\tau(\rho)$ is
attained for so-called connected graph states~\cite{JuMoGu2011pra}.  The
entanglement witness $W_{\rm GHZ}$ for the GHZ state found by the algorithm in
Ref.~\cite{JuMoGu2011prl} is 
\begin{equation}
W_{\textrm{GHZ}}=\mathbbm{1}-\frac{3}{2}\rho_{\textrm{GHZ}}.
\end{equation}
In this case, we have  the same $\opP_s$ for all subsets $s$:
$\opP_s=(\mathbbm{1}-\rho_{\textrm{GHZ}})/2$, with eigenvalues
$0$ and $\tfrac12$ 
($\mathbbm{1}-\opP_s$ has eigenvalues $\tfrac12$ and $1$). For $\opQ_s$ we have 
$\opQ_s= \tfrac12 \, \mathbbm{1}-\rho_{\textrm{GHZ}}^{T_s}$ with eigenvalues
$0$, $\tfrac12$ and $1$ (same eigenvalues for $\mathbbm{1}-\opQ_s$).

We note that $\tau$ is invariant under a change of basis. In our case, this
means that any basis (e.g., a helicity basis) can be used to describe the
polarization vectors~$\epsilon_j$. Furthermore, due to the properties of the
matrix element $M_{\textrm{TC}}$, the entanglement measure $\tau(\rho)$ is
Lorentz invariant as well as gauge invariant in the sense of
Eq.~\eqref{gaugetransformation}.  In principle, $\tau(\rho)$ can be measured by
reconstructing the density matrix~\cite{FaEiMi2007}.  Experimentally, this can
be done by conducting coincidence measurements of the emitted photons with
various polarization filters, as described in Ref.~\cite{JaKwMuWh2001}.

Below, we calculate $\tau(\rho)$ for the same parameter 
values as used in Sec.~\ref{subec:diffcrosssec} 
in the  calculation of 
the differential cross section. It is also interesting to investigate to which extent
the state $\rho$ is mixed. To this end, we have computed, in 
addition to $\tau(\rho)$, the von Neumann entropy 
$Q(\rho)$, defined as~\cite{We1978}
\begin{equation}
\label{defvonNeumennentropy}
\entr(\rho) =-\tr \left(\rho \log_2(\rho)\right) =-\sum_{j=1}^8 u_j\log_2 (u_j) \,,
\end{equation}
where the $u_j$'s  are the eigenvalues of $\rho$.  For any pure state
$\rho=|\psi\rangle\langle\psi|$, we have $\entr=0$ because $\rho^2=\rho$ and
the eigenvalues are $u_j=1$.  A maximally mixed state $\rho_{\textrm{maxmix}}$,
which has equal diagonal entries, and vanishing non-diagonal matrix elements,
has $\entr=(8 \times \tfrac18) \, \log_2 8=3$ in the current case.
$\tau(\rho_{\textrm{maxmix}})=0$, since $\rho_{\textrm{maxmix}}$ can be
factorized as $\rho_{\textrm{maxmix}}=
\mathbbm{1}_{2\times2}\otimes\mathbbm{1}_{2\times2}\otimes\mathbbm{1}_{2\times2}/8$,
and is therefore not entangled.

\begin{figure}
\includegraphics[width=0.5\textwidth]{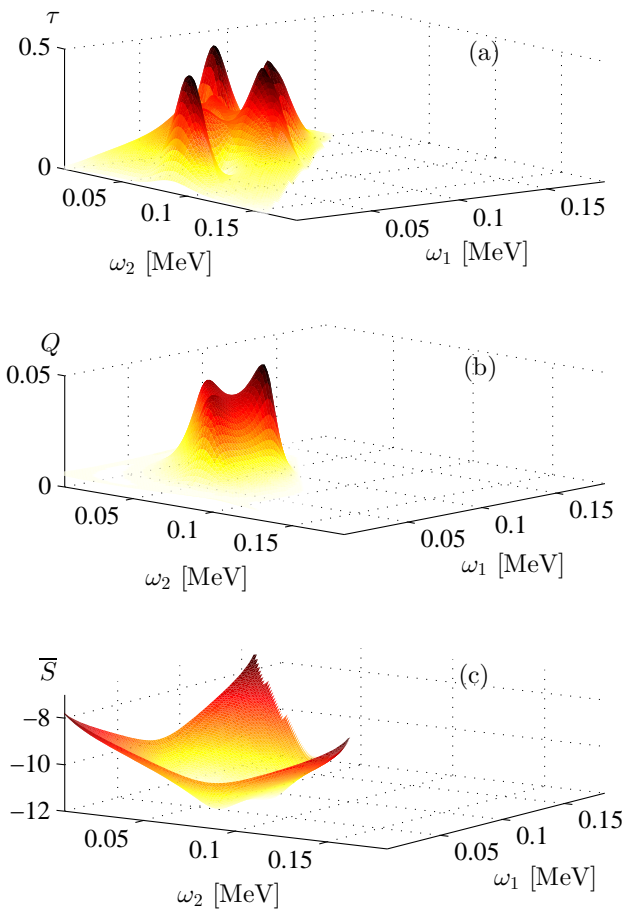}
\caption{\label{fig7} 
\colol We investigate the 
entanglement and von Neumann entropy for the parameters 
given as given in Fig.~\ref{fig3}(a).
Figure~(a) has the entanglement measure $\tau(\rho)$,
whereas in Fig.~(b) we plot the von Neumann entropy $\entr$. 
For completeness, 
we plot in Fig.~(c) the quantity
$\overline{S}$ defined in Eq.~\eqref{defSbar}, i.e.,  the decadic logarithm of 
the differential cross section summed over the 
polarizations of the emitted photons.}
\end{figure}

\begin{figure}
\includegraphics[width=0.5\textwidth]{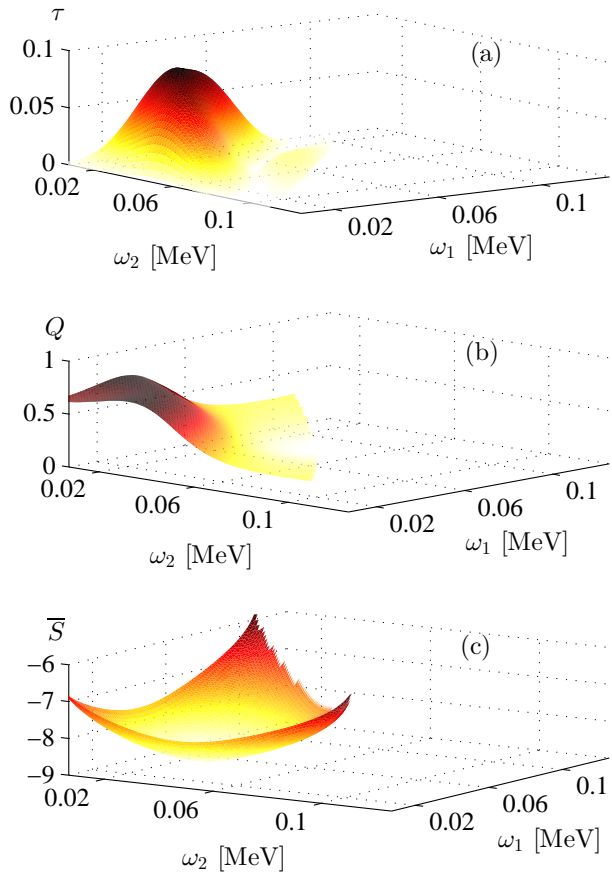}
\caption{\label{fig8} 
\colol We investigate the entanglement of triple Compton photons
for the parameters of Fig.~\ref{fig3}(b), i.e., for a scattering in the
backward cone as seen from the incoming photon.
The entanglement measure $\tau$ and von Neumann entropy $\entr$
are plotted in Figs.~(a) and~(b), respectively.
Finally, in Fig.~(c), we plot
the quantity $\overline{S}$ [see Eq.~\eqref{defSbar} for definition], i.e.,  the decadic logarithm of 
the differential cross section summed over the 
polarizations of the emitted photons. }
\end{figure}

The results of the evaluation of $\tau$ and $\entr$ are shown in
Figs.~\ref{fig7}, and \ref{fig8}, using the parameters of the setup described
in Sec.~\ref{subec:diffcrosssec} (180 keV photons on a stationary target).  For
the calculation of the entanglement measure $\tau(\rho)$, we employ the program
{\sc pptmixer}, made available at \cite{pptmixer} by the authors of
Ref.~\cite{JuMoGu2011prl}.  For comparison, we also show  the differential
cross section summed over the final polarizations, to give an idea about
whether or not the cross section and the entanglement measure $\tau$ are large
at the same parameter values.

By inspecting Figs.~\ref{fig7} and \ref{fig8}, we see that, somewhat
unfortunately, a large value of the entanglement measure $\tau$ is accompanied
by a small value of the differential cross section.  This may limit the
practical usefulness of the three-photon Compton effect as a source of entangled
photons. However, we note that even a small value of $\tau\neq 0$ implies that
the state $\rho$ is genuinely entangled. We also point out that it is natural
for $\tau$, as a measure of correlation, to approach zero for small
$\omega_{1,2,3}$ [at the edges of the ``triangle'' in the $\omega_1\omega_2$
plane where the differential cross section is non-vanishing, see
Figs.~\ref{fig7}(c),~\ref{fig8}(c) and Fig.~\ref{fig3}]. The physical reason
for the lack of entanglement is that in the limit of vanishing $\omega_j$ for
one of the photons, the three-photon Compton process factorizes into a sequential
process of one-photon emission followed by a two-photon event, which leads to a
final state which is not entangled in all three photons. In the extreme case of
two vanishing photon energies (for example, $\omega_1\to 0$ and $\omega_2\to
0$, but finite $\omega_3$), the three-photon Compton process becomes a sequence of
three one-photon events, again with vanishing correlation. 

Finally, we  note that the von Neumann entropy varies depending on the setup. In 
Fig.~\ref{fig7}, we have $Q\ll 1$ in large parts of the $\omega_1$$\omega_2$ plane.
The von Neumann entropy of the state produced in the 
forward cone [see Fig.~\ref{fig7}(b)] is lower than 
the entropy of triplet photon states in the backward cone 
[Fig.~\ref{fig8}(b)]. The entanglement measure $\tau(\rho)$
attains values close to its maximum value $1/2$ in Fig.~\ref{fig7},
indicating that close to  a maximally entangled triplet photon state is produced.
 
We have also computed entangled measures and entropies for the setup described
in Sec.~\ref{subsec:diffcrosssecGeV}, an intense laser beam colliding with a
high-energy electron beam.  The results are very similar to those already
presented in Figs.~\ref{fig7} and \ref{fig8}. Also in this case, a large value
of $\tau$ was only found in angular regions were the cross section is small.
Finally, a limited investigation for the case with $\omega_0=3$ MeV, $E_i=m$
was carried out.  Although potentially difficult to realize experimentally,
this case is interesting since the total cross section for triple Compton
scattering peaks around this value of $\omega_0$ (see Fig.~\ref{fig2}).
However, a  parameter region where both the cross section and $\tau$ are large
could not be found.

%
%
\section{Conclusions}
\label{Sec:Concl}

We have presented a theoretical analysis of the three-photon, or triple Compton effect.
Contrary to the single Compton event, the double and triple processes do not
have a classical analog and therefore, a single low-energy incoming photon is
not sufficient to excite a process with the emission of more than one quantum.
Both the total as well as the differential cross sections for the
double-Compton as well as the triple-Compton processes tend to zero for low
incoming photon energy, as demonstrated in Fig.~\ref{fig2}. The cross section
vanishes also for $\omega_0/m \to\infty$, where we recall that $\omega_0$ is
the energy of the incoming photon. Because the  cross sections for double and
triple scattering vanish for both  $\omega_0 \to 0$ and $\omega_0 \to \infty$,
there has to be a certain initial photon energy for which the double and triple
Compton cross sections have a maximum. With our convention for the infrared
cutoff, this maximum was found to be at $\omega_0\approx 3$~MeV for both the
double and triple Compton effect.

In a previous experiment \cite{MGBr1968}, the triple Compton process was
studied with a symmetric detector geometry, with three detectors oriented in a
``Mercedes--star'' configuration with azimuth angles $\phi_j = 2 j\pi/3$ for
$j=1,2,3$.  The polar angle is assumed to be equal for all three detectors.
The detectors are thus situated at the corners of an equilateral triangle.  We
assume this detector geometry for our cross section calculations (see
Figs.~\ref{fig3} and~\ref{fig5}).  Supplementing a previous
discussion~\cite{LoJe2012}, we consider two example cases for the differential
cross section of the triple Compton process: 180 keV photons at stationary
electrons, and laser photons of energy $2.5$\,eV on GeV electrons.  We find
that the three-photon Compton process is measurable at present synchrotron or laser
facilities, and constitutes one of the most straightforward processes for the
manifestation of high-energy entanglement in the quantum world.  We suggest
that the most favorable experimental setup for measuring the triple Compton
process would be a high-flux synchrotron light source combined with stationary
targets.  Furthermore, as demonstrated in Fig.~\ref{fig7}, a high degree of
entanglement is reached in the final three-photon state, for an arrangement of
the detectors in the forward cone, for the case of an 180\,keV photon impacting
on stationary electrons.

%
%
\begin{acknowledgments} 

This work was supported by the National Science
Foundation (Grant PHY-1068547) and the National Institute of Standards and
Technology (NIST precision measurement grant). E.L. acknowledges partial
support from the FPR program of RIKEN.

\end{acknowledgments}

\end{document}